\documentclass[useAMS,twocolumn]{mn2e}
\usepackage{graphicx}
\usepackage{times}
\usepackage{amsmath}
\usepackage{epstopdf}
\usepackage[authoryear]{natbib}

\begin{document}

\title[Signatures of SNIa Progenitors]
{Observational Signatures of SNIa Progenitors, as Predicted by Models}
\author[Hillman et al.]
{Y.~Hillman,$^1$ D.~Prialnik,$^1$ A.~Kovetz,$^{1,2}$ and M.~M.~Shara$^3$\\
$^1${Department of Geophysics and Planetary Sciences, Raymond and Beverly Sackler Faculty of Exact Sciences, Tel-Aviv University, Tel-Aviv 69978, Israel.}\\
$^2${School of Physics and Astronomy, Raymond and Beverly Sackler Faculty of Exact Sciences, Tel Aviv University, Tel-Aviv 69978, Israel.}\\
$^3${Department of Astrophysics, American Museum of Natural History, Central Park West and 79th street, New York, NY 10024-5192, USA.}\\
}

\maketitle
\begin{abstract}
A definitive determination of the progenitors of type Ia supernovae (SNIa) has been a conundrum for decades. The single degenerate scenario --- a white dwarf (WD) in a semi-detached binary system accreting mass from its secondary --- is a plausible path; however, no simulation to date has shown that such an outcome is possible. In this study, we allowed a WD with a near Chandrasekhar mass of $1.4M_\odot$ to evolve over tens of thousands of nova cycles, accumulating mass secularly while undergoing periodic nova eruptions. 
We present the mass accretion limits within which a SNIa can possibly occur. 
The results showed, for each parameter combination within the permitted limits, tens of thousands of virtually identical nova cycles where the accreted mass exceeded the ejected mass, i.e. the WD grew slowly but steadily in mass.  Finally, the WD became unstable, the maximal temperature rose by nearly two orders of magnitude, heavy element production was enhanced by orders of magnitude and the nuclear and neutrino luminosities became enormous. We also found that this mechanism leading to WD collapse is robust, with WDs in the range $1.0$ \textendash $1.38M_\odot$, and an accretion rate of $5\times10^{-7}M_\odot yr^{-1}$, all growing steadily in mass. These simulations of the onset of a SNIa event make observationally testable predictions about the light curves of pre-SN stars, and about the chemistry of SNIa ejecta. 
\end{abstract}


\section{Introduction}
 
Novae occur in semi-detached binary systems comprising a WD accreting hydrogen-rich matter via an accretion disk from a red dwarf (RD) or red giant (RG) companion. They are powered by thermonuclear runaways (TNR) in their WD's hydrogen-rich envelopes \cite[]{Starrfield1972}. After ejecting their erupting envelopes, novae self-extinguish \cite[]{Prialnik1978}.\par
Early nova simulations exclusively dealt with the WD, and most of them only with the accretion phase, up to the TNR. The first full-cycle nova simulation was carried out by \cite{Prialnik1986} on a main-frame computer. With the progression of technology, and the appearance of PCs with high speed and large memory, simulations of several, consecutive cycles were carried out \cite[]{Shara1993, Prikov1995}. It became clear \cite[]{Prialnik1995} that a simulation of a nova on a WD of given chemical composition (He, CO or ONe) depended on three parameters: the WD mass $M_{\rm WD}$, its core temperature $T_{c}$ and the accretion rate $\dot M$. Grids of nova simulations for wide ranges of these three parameters were calculated \cite[]{Prikov1995, Yaron2005} as well as simulations of long-term evolution, involving {\sl thousands} of cycles, during which the WD's mass secularly \textsl{decreased} due to the difference between the accreted and the ejected masses \cite[]{Epelstain2007, Idan2013}.\par 
There are two primary scenarios that can lead to a SNIa: the single degenerate (SD) and the double degenerate (DD). The former is the case of a WD accreting hydrogen rich material from its secondary, less massive star, undergoing periodic nova or nova-like flashes while retaining mass, and finally exceeding the Chandrasekhar limiting mass \cite[]{Whelan1973, Hillebrandt2000}. The latter is the case of the merging of two WDs in a close binary, by spiraling into each other due to angular momentum loss, while their combined mass exceeds the Chandrasekhar mass \cite[]{Iben1984, Webbink1984}. In addition there is the sub-Chandrasekhar scenario involving helium detonation on intermediate mass WDs which, by satisfying certain conditions, can cause a shock wave into the core and ignite it \cite[]{Livne1990}.   
The SD scenario is believed to be the final outcome of a recurrent nova producing system in which the WD's mass secularly \textsl{increases}. If a WD in a binary system can somehow be made to secularly accrete enough matter (over thousands of nova cycles) to exceed the Chandrasekhar mass, it will fuse carbon and explode as a type Ia supernova (SNIa).
To test this hypothesis, we have attempted to {\textquotedblleft}{push}{\textquotedblright} a sub-Chandrasekhar mass WD, in a semi-detached binary system, over the mass required to initiate a SNIa. We test a range of accretion rates that lead to net mass accumulation on the surface of the WD, through a long series of nova cycles. Our key goal is to predict the observational signature of a pre-SN that might be detectable observationally.

In the next section we describe the model and the parameters used. In Section \ref{sec_results} we present the results of long-term evolutionary calculations.   
In Section \ref{:Discussion} we discuss the limits and limitations of these simulations and present three supersoft X-ray sources that display many of the characteristics of the models of Section \ref{sec_results}. Our conclusions follow in Section \ref{:concl}. 

\section{Model \& Calculations Description}\label{:descrip}
\begin{figure*}
\begin{center}
{\includegraphics[viewport = 30 23 748 583,clip,width=0.99\columnwidth]{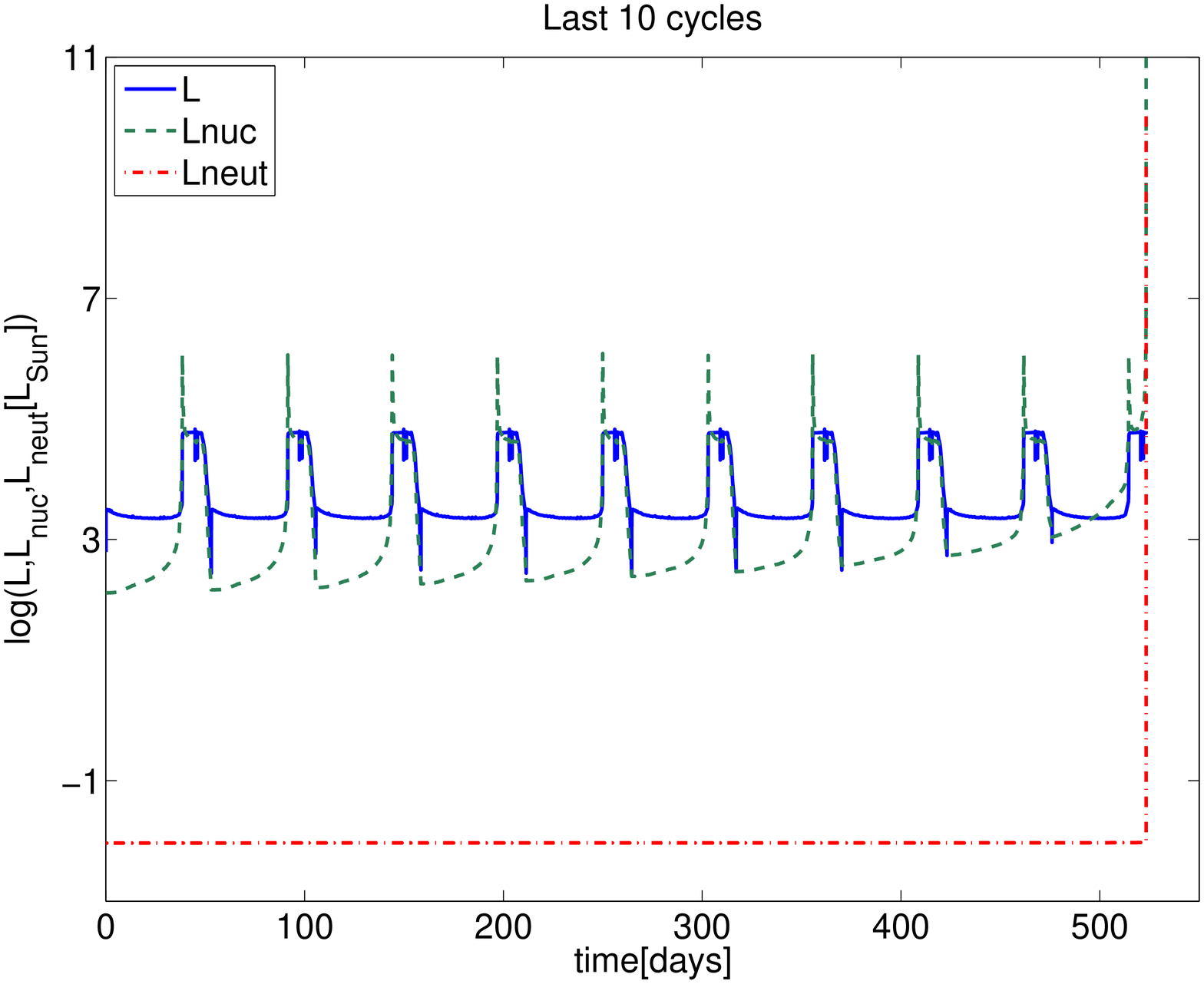} }
{\includegraphics[viewport = 24 19 746 579,clip,width=0.99\columnwidth]{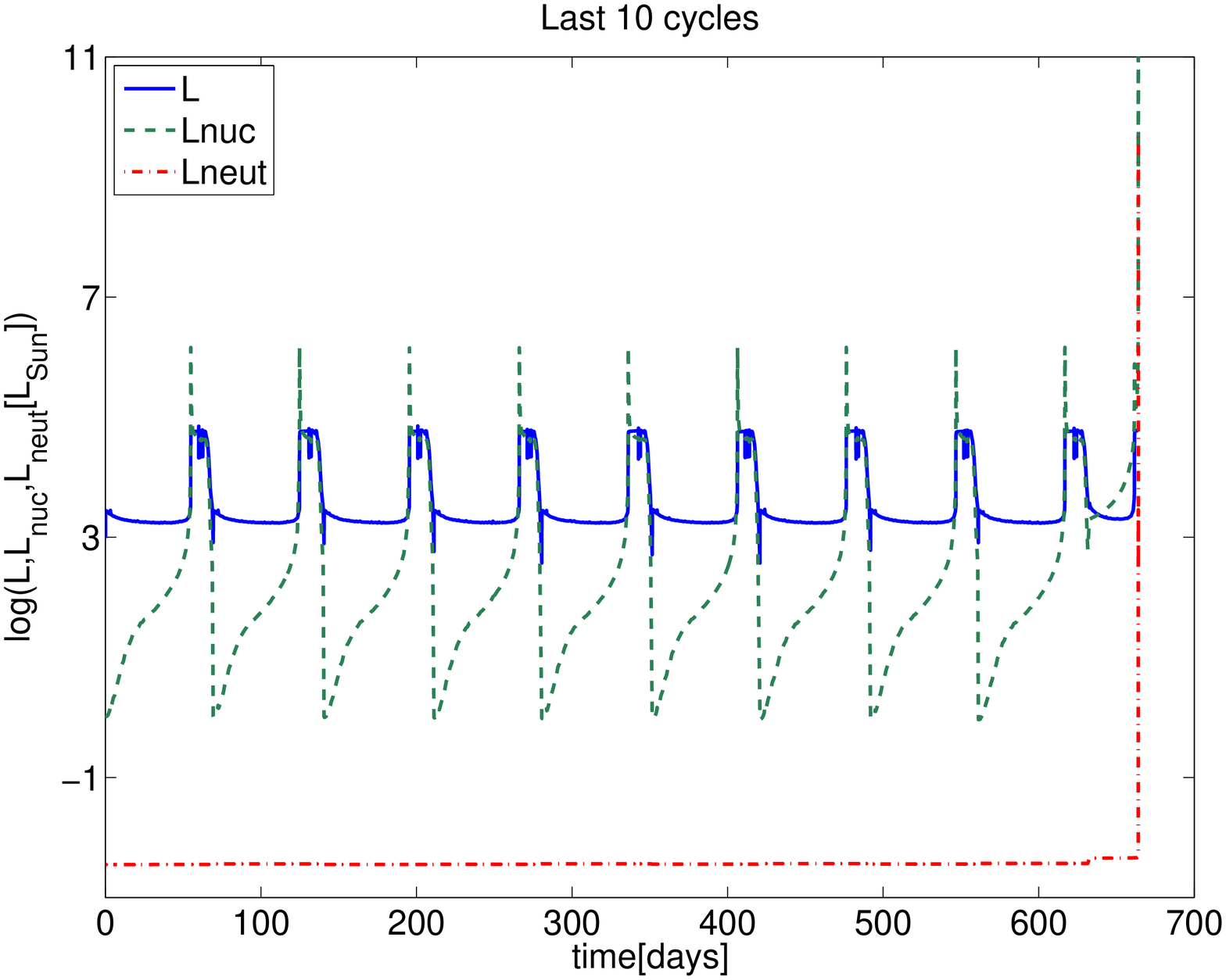} }
\caption{\label{fig:10lastcycs}Evolution of luminosities --– bolometric, nuclear and neutrino for the last ten cycles. Left: for an accretion rate of $5\times 10^{-7}M_{\odot}yr^{-1}$; right: for an accretion rate of $4\times 10^{-7}M_{\odot}yr^{-1}$.} 
\end{center}
\end{figure*}

Throughout the calculations, we use a hydrodynamic Lagrangian code, designed to evolve a WD through multiple consecutive nova cycles \cite[]{Prikov1995}. The code simulates the evolution of a WD accreting mass and periodically producing nova outbursts. It includes OPAL opacities, convection according to the mixing-length theory, diffusion for all elements, accretional heating, a mass-loss algorithm that applies a steady, optically-thick supersonic wind solution and a nuclear reaction network, including $\rm{H}$, $^{3}\rm{He}$, $^{4}\rm{He}$, neutrons and 40 heavy element isotopes, namely; $^{12}\rm{C}$, $^{13}\rm{C}$, $^{14}\rm{C}$, $^{13}\rm{N}$, $^{14}\rm{N}$, $^{15}\rm{N}$, $^{14}\rm{O}$, $^{15}\rm{O}$, $^{16}\rm{O}$, $^{17}\rm{O}$, $^{18}\rm{O}$, $^{17}\rm{F}$, $^{18}\rm{F}$, $^{19}\rm{F}$, $^{18}\rm{Ne}$, $^{19}\rm{Ne}$, $^{20}\rm{Ne}$, $^{21}\rm{Ne}$, $^{22}\rm{Ne}$, $^{20}\rm{Na}$, $^{21}\rm{Na}$, $^{22}\rm{Na}$, $^{23}\rm{Na}$, $^{22}\rm{Mg}$, $^{23}\rm{Mg}$, $^{24}\rm{Mg}$, $^{25}\rm{Mg}$, $^{26}\rm{Mg}$, $^{24}\rm{Al}$, $^{25}\rm{Al}$, $^{26}\rm{Al}$, $^{27}\rm{Al}$, $^{27}\rm{Si}$, $^{28}\rm{Si}$, $^{29}\rm{Si}$, $^{30}\rm{Si}$, $^{28}\rm{P}$, $^{29}\rm{P}$, $^{30}\rm{P}$ and $^{31}\rm{P}$, interacting through $\sim 200$ nuclear reactions. The dynamical phases are calculated by solving the equation of motion along with the energy balance equation, rather than imposing hydrostatic equilibrium. Mass loss is calculated continually, according to the mass loss rate derived from the optically thick wind solution. Each evolution run requires three initial parameters; the WD mass ($M_{\rm WD}$), the core temperature ($T_c$) and the accretion rate ($\dot{M}$). The initial chemical composition is set as well, both for the WD and for the accreted material. 

In order to result in a SNIa explosion, a WD must reach the Chandrasekhar limit \cite[]{Chandrasekhar1931}. We must numerically follow many consecutive nova cycles while the WD secularly accumulates mass. To allow for a continually growing WD, while keeping the numerical grid no larger in size than necessary for accurate and reproducible calculations \cite[]{Epelstain2007}, we have modified the code to evolve nova models through tens of thousands of cycles without intervention (the adaptive grid size never needed to exceed a few hundred shells).

Full evolutionary calculations were carried out for a WD composed of carbon and oxygen in equal mass fractions, with an initial mass of $1.4M_{\odot}$ and an initial core temperature of $3\times 10^7{\rm K}$, accreting solar-composition material at a constant rate. We sampled accretion rates in the range $10^{-8}$\textendash $10^{-6} M_{\odot}$~yr$^{-1}$, specifically: $0.1$, $0.2$, $0.3$, $0.4$, $0.5$, $1.0$, $2.0$, $4.0$, $5.0$, $6.0$, $7.0$, and $8.0\times 10^{-7}M_{\odot}$~yr$^{-1}$. Initial core temperatures of 1 and $5\times 10^7{\rm K}$ were explored as well. 
Cases with accretion rates of $7.0\times 10^{-7}M_{\odot}yr^{-1}$ and higher were not able to withstand the rapid mass accretion and developed super-Eddington luminosities that prevented mass accumulation. 
Accretion rates of $0.2\times 10^{-7}M_{\odot}yr^{-1}$ and lower led to secular mass loss \cite[]{Yaron2005, Idan2013} and therefore cannot be SNIa candidates. 

Models with accretion rates between $0.3$ and $6.0\times 10^{-7}M_{\odot}yr^{-1}$ were found to steadily grow in mass.
These mass accretion rates are broadly consistent with mass loss and transfer rates in symbiotic binaries.
Within these limits, and for accretion rates of $0.5$, $4.0$, $5.0$ and $6.0\times 10^{-7}M_{\odot}yr^{-1}$, the models ran for $\sim55,000$, $\sim39,000$, $\sim36,000$ and $\sim10,000$ consecutive nova outbursts, gaining a net total mass of $\sim13.3$, $\sim2.3$, $\sim2.9$ and $\sim0.96\times 10^{-4}M_{\odot}$, respectively, until thermonuclear instability developed and a gigantic explosion ensued, resembling the onset of a supernova. 

\section{Results}\label{sec_results}
None of our models developed steady hydrogen burning, regardless of the high accretion rate. It has long been argued that steady burning cannot arise in WDs accreting hydrogen-rich material, and both this paper and recent works have confirmed this conclusion, e.g., \cite{Starrfield2012} and \cite{Idan2013}. 
All claims that steady burning can exist, e.g., \cite{Hachisu2012}, are based on analytic models that cannot calculate unstable nuclear burning. This unstable burning underlies the flashes and net mass accretion we are about to describe.  
All models went through a long series of very similar --- almost identical --- cycles of accretion, luminosity rise, expansion, contraction and luminosity drop.  

We focus on the results for an initial core temperature of $3\times 10^7{\rm K}$, although we have also tested lower and higher temperatures. The lower core temperature of $1\times 10^7{\rm K}$ showed a slightly shorter eruptive cycle duration (i.e., the time between successive eruptive events, or recurrence period) and the higher core temperature of $5\times 10^7{\rm K}$ showed a slightly longer cycle duration (in both cases, about a $5\%$ difference compared with the corresponding baseline model). The shorter and longer cycle durations affected the accreted and ejected mass accordingly, however, the net mass gain was essentially the same for the three temperatures. Other evolutionary features showed no significant variations between the different initial core temperatures.

Our code is not designed to follow the entire evolution of a WD and to simulate it experiencing a SN explosion. Rather, it is designed, and is sufficiently accurate to follow the accreting and periodically erupting pre-SN WD envelope, and to detect the onset of a SN explosion. This explosion --- referred to as the final outburst --- is evident, with peak temperatures reaching a few $10^9{\rm K}$, huge nuclear and neutrino luminosities, and fast nuclear burning up to the heaviest elements included in our network.

\begin{figure}
\begin{center}
{\includegraphics[viewport = 50 45 757 590,clip,width=0.99\columnwidth]{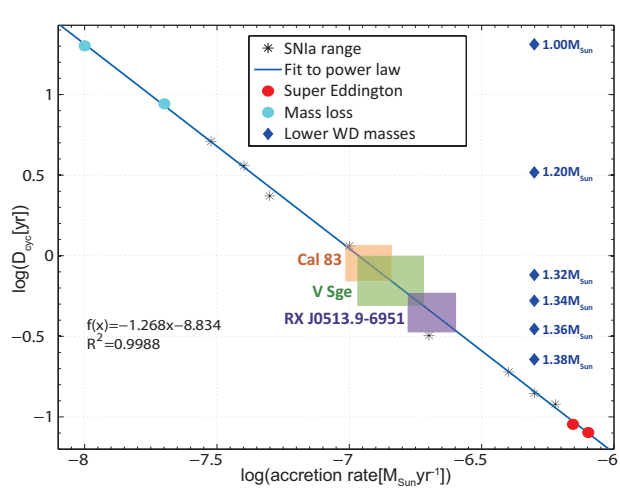}}
\caption{\label{fig:P(Mdot)}Cycle duration vs. accretion rate for a $1.40M_\odot$. The blue diamonds represent the cycle duration for an accretion rate of $5\times 10^{-7}M_\odot yr^{-1}$ for the lower mass WDs which is discussed in \S\ref{:SNIa}} 
\end{center}
\end{figure}

 Fig.\ref{fig:10lastcycs} shows the last ten nova flashes leading to the final outburst for the accretion rates   $4.0$ and $5.0\times 10^{-7}M_{\odot}yr^{-1}$. The final explosion, appearing as a 16 orders of magnitude increase in nuclear luminosity, occurs in a matter of minutes. For the $5.0\times 10^{-7}M_{\odot}yr^{-1}$ case this lasted $\sim14$ minutes, while this phase is shorter for higher accretion rates (e.g. $\sim7$ minutes for $6.0\times 10^{-7}M_{\odot}yr^{-1}$) and longer for lower accretion rates (e.g. $\sim33$ minutes for $4.0\times 10^{-7}M_{\odot}yr^{-1}$). The duration of each nova cycle is almost the same throughout the evolution of each model: shorter for higher accretion rates and longer for lower accretion rates. Fig.\ref{fig:10lastcycs} shows a cycle duration of $\sim50$ days for an accretion rate of $5.0\times 10^{-7}M_{\odot}yr^{-1}$, and of $\sim65$ days for an accretion rate of $4.0\times 10^{-7}M_{\odot}yr^{-1}$. This behavior is further demonstrated in Fig.\ref{fig:P(Mdot)}, on a double logarithmic scale, where the correlation between cycle duration in years, $D$, and accretion rate in solar masses per year, may be approximated remarkably well by a simple fit formula: 
\begin{equation}\label{eqPMdot}
\log(D)=-1.27\times \log(\dot{M})-8.83.
\end{equation}

\cite{Kovetz1994} derived the durations of $\sim1$ year, and $\sim20$ years for accretion rates of $10^{-7}M_\odot yr^{-1}$ and $10^{-8}M_\odot yr^{-1}$ respectively, and \cite{Walder2008} find via simulation that the observed cycle duration of $\sim22$ years for the recurrent nova RS-Oph is in agreement with an accretion rate of $10^{-8}M_\odot yr^{-1}$. Both derived timescales agree well with Eq.\ref{eqPMdot}, as do the simulations performed by \cite{Yaron2005}.

The observed limits of $D$ for three accreting objects in binary systems --- Cal 83, V Sge and RX J0513.9-6951 (presumed to be massive WDs) --- are presented in Fig.\ref{fig:P(Mdot)} as well. We used the cycle durations of these objects in our power law to derive limits to their accretion rates, yielding an approximated area in which these WD's could exist. These systems are discussed further in \S\ref{:conclusions}.

\subsection{Regular Nova Cycles}\label{:reg_cycs}

\begin{figure}
\begin{center}
{\includegraphics[viewport = 45 15 752 560,clip,width=0.99\columnwidth]{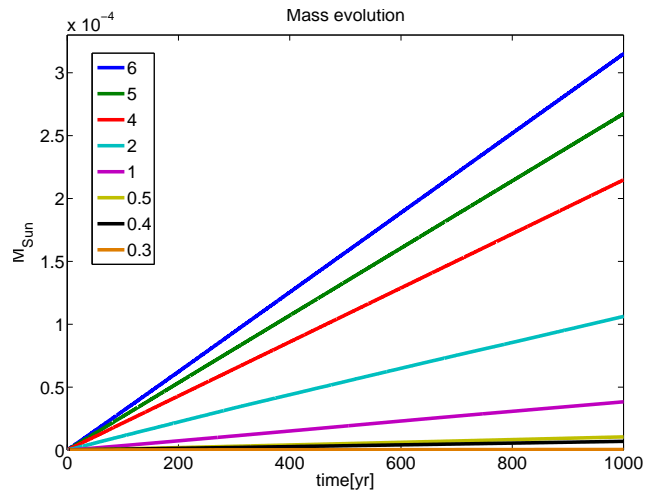}}
\caption{\label{fig:Mass_evol}Mass growth over time for a $1.4M_\odot$ for the accretion rates; $0.3$, $0.4$, $0.5$, $1.0$, $2.0$,$4.0$,$5.0$ and $6.0\times 10^{-7}M_\odot yr^{-1}$} 
\end{center}
\end{figure}

The tens of thousands of nova cycles, leading to the onset of the supernova explosion, demonstrate virtually identical behaviors. Changes between eruptions are very small, slow, secular and almost linear. The accreted and ejected masses are almost identical at each cycle. The accreted mass exceeds the ejected mass by $\sim0.7-2.4\times 10^{-8}M_{\odot}$ per cycle, depending on the mass accretion rate. The mass evolution over time is plotted in Fig.\ref{fig:Mass_evol} for the accretion rates focused on in this study. It clearly shows that for higher accretion rates, more mass will be retained.\par
The duration of each cycle ranges from $\sim45$ days for the extremely fast accretion rate of $6\times 10^{-7}{M}_{\odot}yr^{-1}$ to $\sim8.75$ years for the borderline case $0.2\times 10^{-7}{M}_{\odot}yr^{-1}$, which secularly maintains a constant mass (see Fig.\ref{fig:P(Mdot)}). The maximum WD shell flash temperature ranges from $\sim1.75$ to $\sim2.44\times10^8{\rm K}$ for $6$ to $0.2\times 10^{-7}{M}_{\odot}yr^{-1}$, respectively. The central temperature ($T_c$) grows steadily, but slightly, about $1-3\%$. The central density ($\rm Rho_c$), initially $\sim9\times10^9g/cm^3$, grows by about $2-8\%$, over the entire evolution depending on the model.

The composition of the ejecta shows no difference between cycles.
The abundances of the envelope composition, for a regular cycle and for the explosive cycle of the $5\times 10^{-7}{M}_{\odot}yr^{-1}$ model are given in Table \ref{Tab:envelopecomposition} and discussed in \S\ref{:SNIa}.

\begin{table}
\begin{center}
\begin{tabular}{|c|c|c|}
\hline
{Element} & {Regular cycle }&{Explosive cycle }\\ 
\hline
{H} &{8.10 (-05)} &{4.51 (-05)}\\
{n} &{1.41 (-21)} &{3.20 (-05)}\\
{He4} &{9.74 (-01)} &{8.52 (-01)}\\
{C12} &{5.15 (-03)} &{7.18 (-03)}\\
{C13} &{1.55 (-05)} &{1.83 (-06)}\\
{C14} &{4.86 (-04)} &{3.95 (-04)}\\
{N13} &{1.24 (-07)} &{2.23 (-07)}\\
{N14} &{1.54 (-02)} &{1.09 (-02)}\\
{N15} &{1.37 (-04)} &{1.26 (-04)}\\
{O16} &{3.47 (-03)} &{4.85 (-03)}\\
{O17} &{1.23 (-05)} &{1.01 (-05)}\\
{O18} &{8.70 (-04)} &{6.15 (-04)}\\
{F18} &{1.03 (-09)} &{4.85 (-04)}\\
{F19} &{3.31 (-06)} &{7.57 (-06)}\\
{Ne20} &{2.58 (-08)} &{2.70 (-05)}\\
{Ne21} &{4.47 (-10)} &{1.83 (-04)}\\
{Ne22} &{2.57 (-08)} &{1.87 (-04)}\\
{Na22} &{4.3 (-15)} &{6.38 (-07)}\\
{Na23} &{6.6 (-11)} &{2.18 (-05)}\\
{Mg24} &{3.56 (-12)} &{1.96 (-04)}\\
{Mg25} &{5.25 (-10)} &{4.45 (-04)}\\
{Mg26} &{4.40 (-10)} &{3.11 (-05)}\\
{Al26} &{5.62 (-13)} &{1.71 (-06)}\\
{Al27} &{4.72 (-11)} &{4.77 (-04)}\\
{Si28} &{2.60 (-09)} &{1.20 (-01)}\\
{Si29} &{1.28 (-11)} &{1.20 (-03)}\\
{Si30} &{5.00 (-11)} &{1.98 (-06)}\\
{P31} &{0.00 (-00)} &{6.05 (-06)}\\
\hline
\end{tabular}
\end{center}
\caption{\label{Tab:envelopecomposition}Composition (in mass fraction) of the WD's envelope for a regular nova cycle and for the last cycle from the $5\times 10^{-7}M_{\odot}yr^{-1}$ accretion rate model.}
\end{table}

\begin{figure}
\begin{center}
{\includegraphics[viewport = 45 15 715 548,clip,width=0.99\columnwidth] {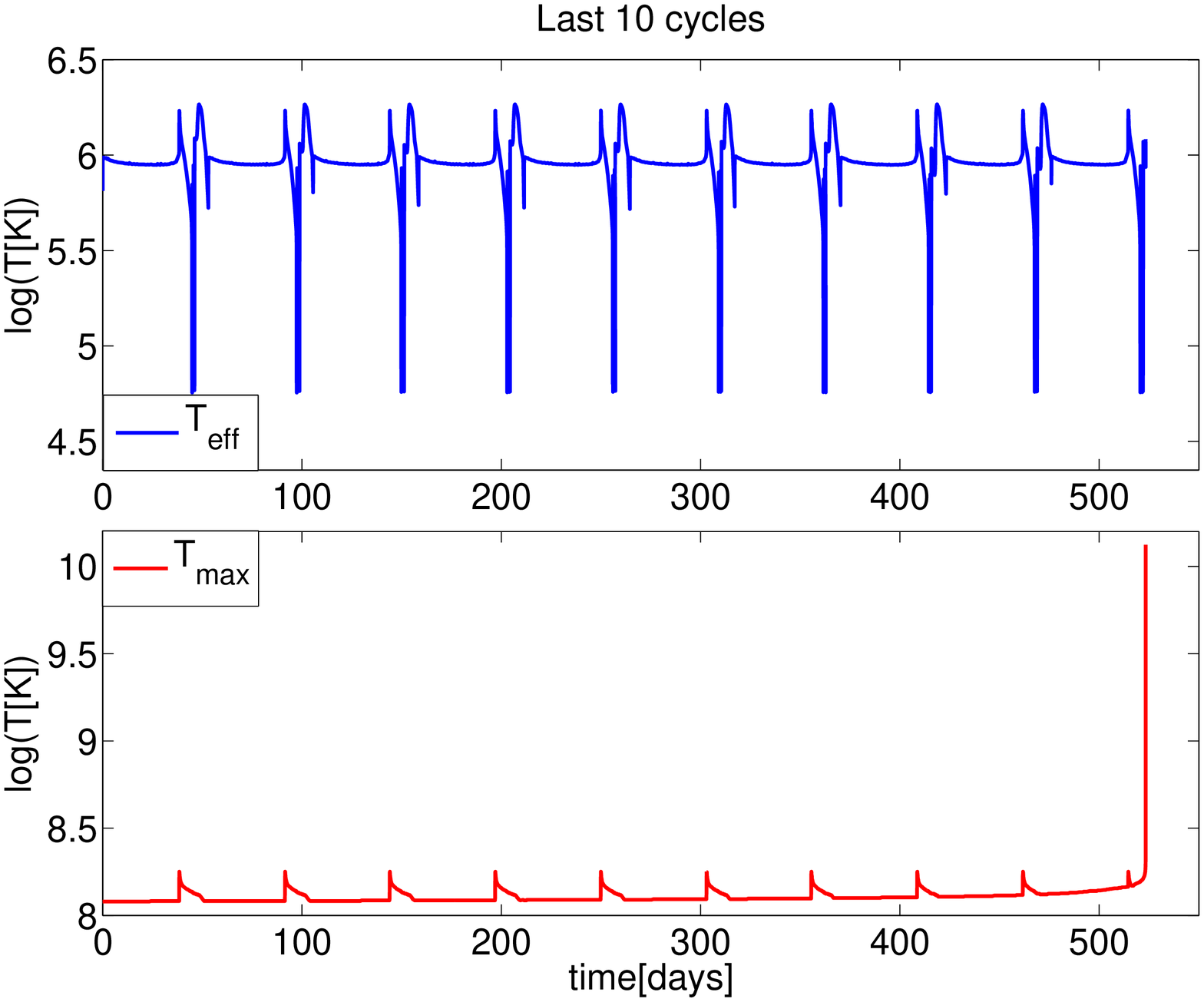}}
\caption{\label{fig:TeffTmax} Temperature change over the last ten cycles for the $5\times 10^{-7}M_{\odot}yr^{-1}$ accretion rate model. Top: $T_{\rm eff}$; bottom: $T_{\rm max}$.} 
\end{center}
\end{figure}

\begin{figure}
\begin{center}
{\includegraphics[viewport = 40 20 747 575,clip,width=0.99\columnwidth] {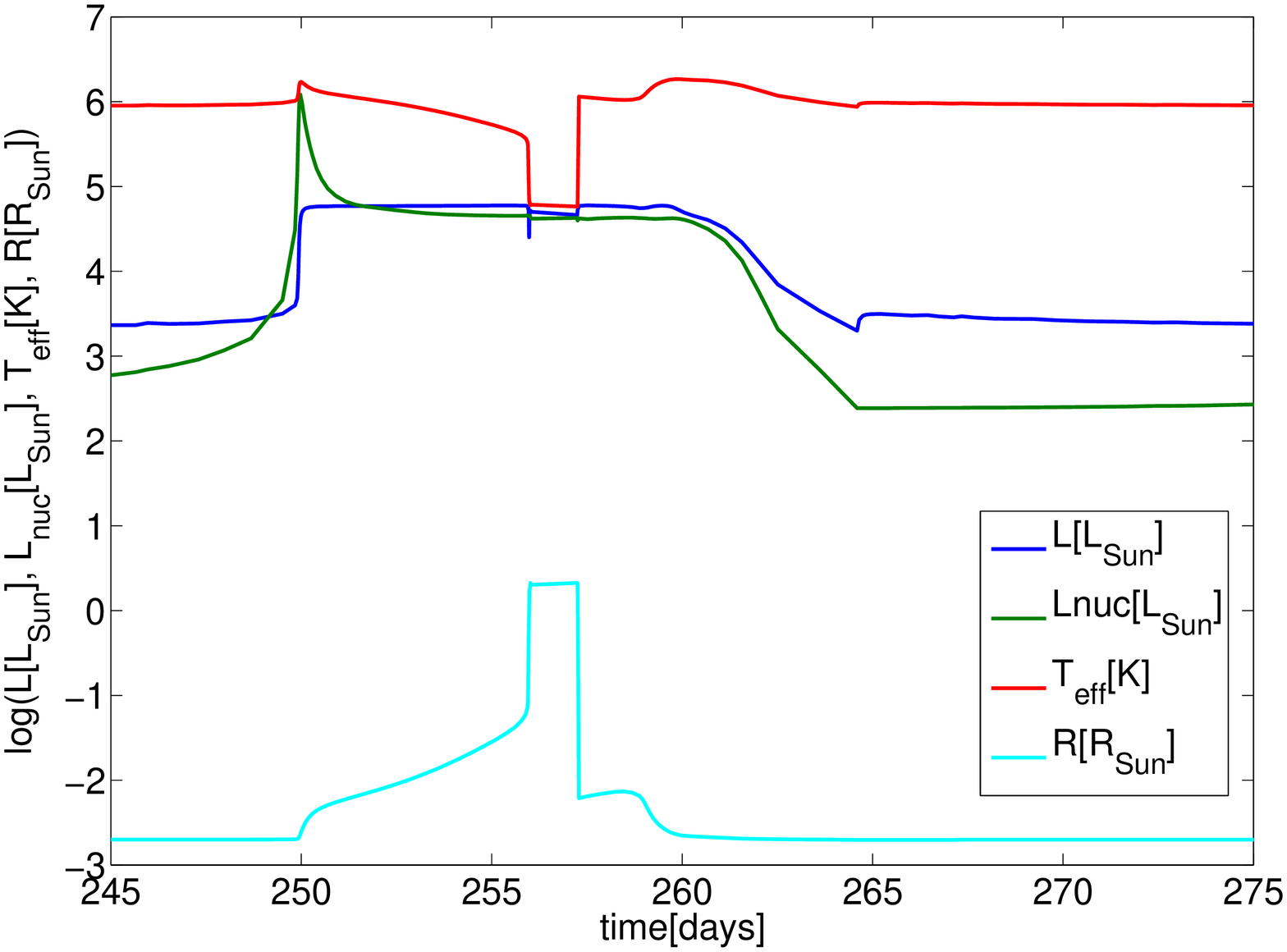}}
\caption{\label{fig:ZOOMLTR} Bolometric and nuclear luminosities, effective temperature and radius over one full cycle from the $5\times 10^{-7}M_{\odot}yr^{-1}$ accretion rate model.} 
\end{center}
\end{figure}

Each nova cycle begins with a pre-nova flash where the effective temperature rises rapidly and then drops (Fig.\ref{fig:TeffTmax}, top panel and a close up of a single eruptive cycle in Fig.\ref{fig:ZOOMLTR}). This is typical of any nova outburst \cite[]{Prialnik2001, Hillman2013}. The drop in effective temperature is due to envelope expansion caused by the nuclear flash, and lasts through the phase of mass ejection. When mass loss ceases, contraction ensues, the effective temperature rises once again and then slowly declines. This is demonstrated in Fig.\ref{fig:ZOOMLTR}, where the radius expands as the luminosities (nuclear and bolometric) rise and contracts  as the luminosities decline. During the two effective temperature peaks, the WD radiates mostly in the UV and the X-ray passbands, while in between --- during the nova phase --- it is mostly detectable in the visible passband \cite[fig.3]{Hillman2013}. In massive, secular mass-gaining WDs ($1.4M_{\odot}$), a typical nova cycle lasts between a couple of months and a few years (see Fig.\ref{fig:P(Mdot)}) within which both peaks occur. 
Notice the large change of $\sim1.5$ orders of magnitude in the effective temperature throughout a single cycle (Fig.\ref{fig:TeffTmax} and \ref{fig:ZOOMLTR}), which repeats itself every cycle. This variability can be observationally searched for and used as an additional tool for identifying nova binaries on the way to a SN explosion.

The maximal WD temperature variation is shown in the lower panel of Fig.\ref{fig:TeffTmax}. We note that between outbursts it stays above $10^8{\rm K}$, meaning that nuclear burning continues: helium is fusing, but more slowly than the rate at which it is produced.

\subsection{Final Explosive Cycle}\label{:SNIa}
\begin{figure}
\begin{center}
{\includegraphics[viewport = 0 0 800 555,clip,width=0.99\columnwidth] {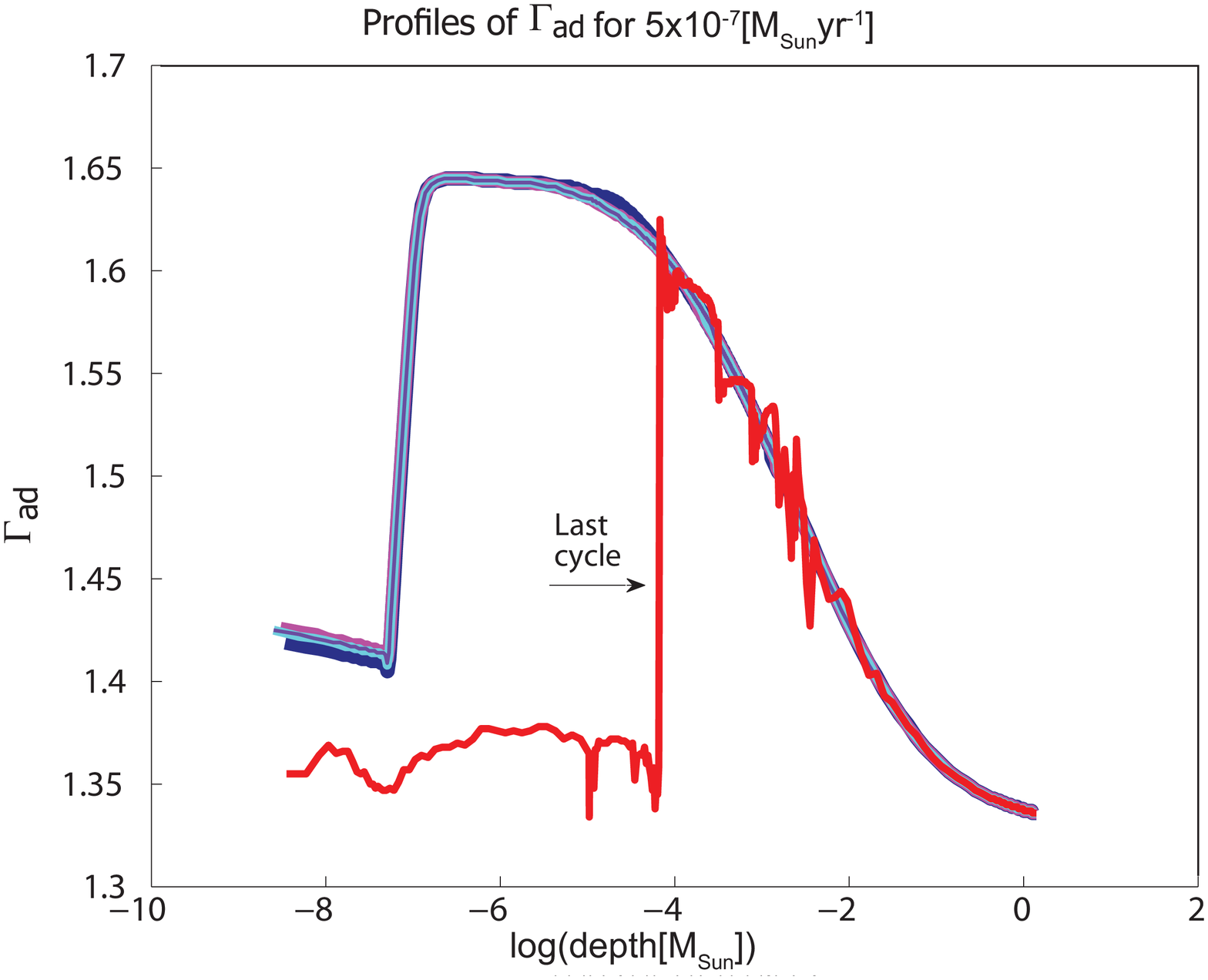}}
{\includegraphics[viewport = 50 30 765 570,clip,width=0.99\columnwidth] {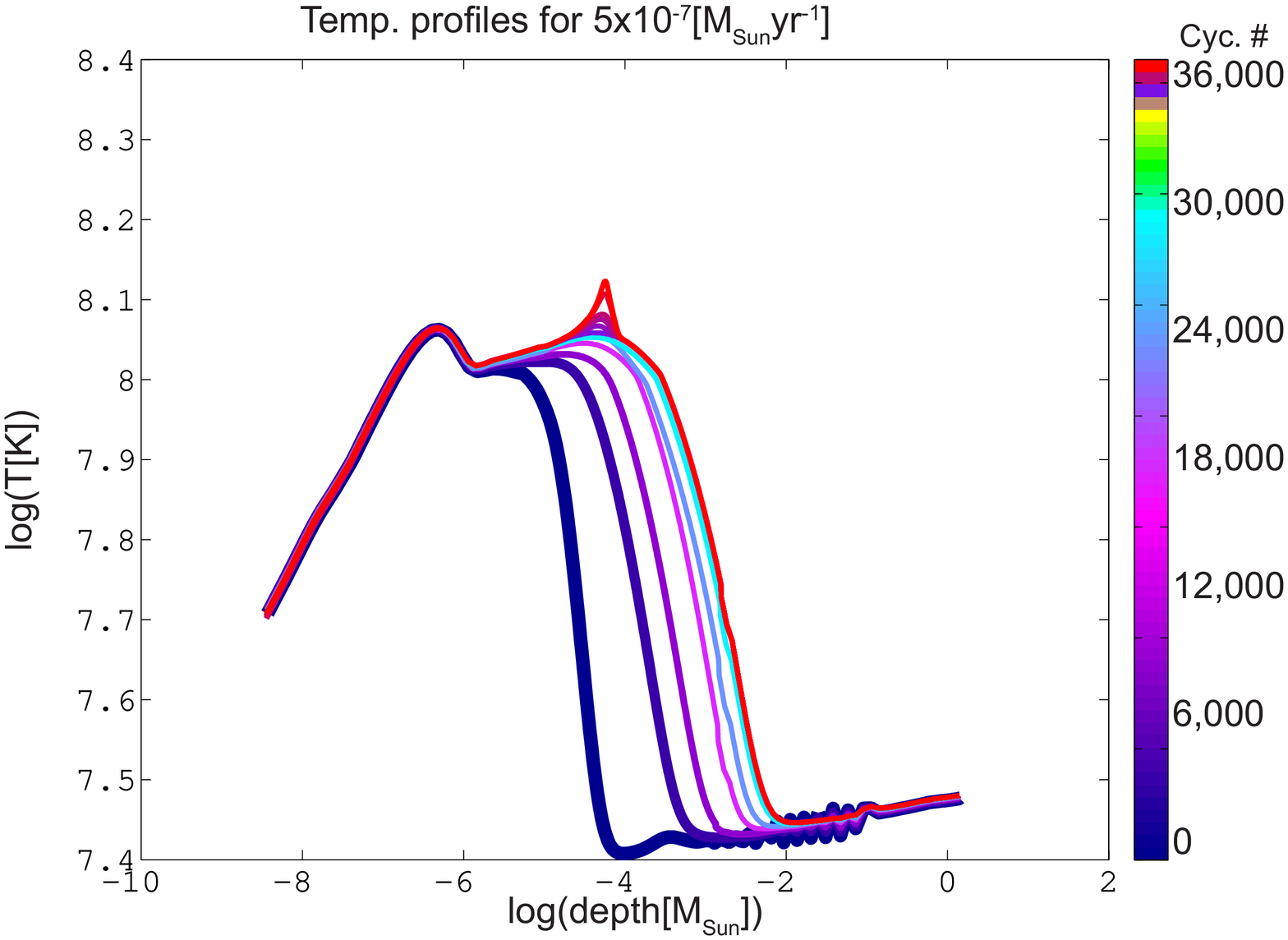}}
\caption{\label{fig:profiles} Progression in time for the $5\times 10^{-7}M_{\odot}yr^{-1}$ accretion rate model for the adiabatic exponent profile (top) showing the drastic change in the profile of the last cycle, compared with earlier cycles which fall upon the same line, and for the temperature profile (bottom) where the heat slowly deepens with time.} 
\end{center}
\end{figure}

A violent instability occurs during the accretion of the final cycle. $T_{max}$ rapidly rises to a few $10^9{\rm K}$, as shown in the bottom panel of Fig.\ref{fig:TeffTmax}, the nuclear and neutrino luminosities rise by $\sim10$ orders of magnitude in just seconds, as shown in Fig.\ref{fig:10lastcycs}, and
there is fast nuclear burning in the envelope into the heaviest elements included in our network --- the abundances of everything heavier than oxygen  increases to a mass fraction of at least five orders of magnitude higher than in a regular cycle. The neutron abundance increases more than 16 orders of magnitude and becomes comparable to the heavy element abundance (Table \ref{Tab:envelopecomposition}). To illustrate the dynamical instability, we plot in Fig.\ref{fig:profiles} (top panel) the adiabatic exponent, $\Gamma_{ad}$, noting that at the final cycle, it drops to 4/3 throughout the outer layers. 
The temperature profile shows the heat gradually penetrating, throughout the evolution, into the depth of the WD (Fig.\ref{fig:profiles}, bottom panel).\par

During the final cycle, close to the time of the final explosion, the temperature profile undergoes great changes. The temperature peaks at well above $10^9{\rm K}$ at a depth of $\sim 10^{-2}M_{\odot}$ --- far below the accreted envelope. 
It begins with the development of a steep rise within the envelope, at a depth of $\sim5\times 10^{-5}M_{\odot}$, where the temperature reaches as high as $10^9{\rm K}$, boosting the production of heavy elements. At this point, the density, which grew smoothly with the depth, drops about two orders of magnitude throughout the envelope, indicating rapid expansion. As the temperature continues to rise and the nuclear burning front advances inward, an even larger temperature peak develops deeper in the WD and finally, the temperature rises at the WD center as well.

The composition profile changes immensely during this time. Before the onset of the explosive instability, the core is composed mainly of $^{12}$C and $^{16}$O, the envelope is composed almost completely of $^4$He with $\sim10^{-2}$ of $^{14}$N, $\sim10^{-9}$ of $^{28}$Si and very small amounts of some other heavy elements, and the outer layers of the envelope are composed of accreted $^4$He and H. At the onset of the final runaway, the profile is considerably different: the outer layers are the same as before, but about half the $^4$He in the rest of the envelope has been burnt into $^{28}$Si. Many other heavy elements, such as isotopes of Al, Si and P, which were either trace or nonexistent before, grow in abundance by many orders of magnitude. The core remains composed mainly of C and O, but at the depth where the temperature rises to a few times $10^9{\rm K}$ there is intense burning that produces $^{28}$Si and $^{31}$P locally. As mentioned before, the code includes heavy elements only up to $^{31}$P. Examination of Table \ref{Tab:envelopecomposition} reveals a huge amount of neutron production as well. This mix of heavy elements and neutrons at this extreme temperature would undoubtedly continue to fuse into heavier elements, such as nickel, cobalt etc., enhancing the instability and leading eventually to a powerful explosion --- a SNIa.   
 
\section{Discussion}\label{:Discussion}
\subsection{The Chandrasekhar limit }\label{:Chandrasekhar}
The Chandrasekhar limit corresponding to a composition of C+O, that is, $\mu_e=2$, is $1.46M_{\odot}$ for the most recent values of physical constants. However, this value is based on the assumption of complete electron degeneracy, without any interactions. \cite{Hamada1961} calculated more accurate critical mass values back in 1961. 
They used \citeauthor{Salpeter1961}'s (1961) zero-temperature equation of state, which takes into account many corrections, such as Coulomb interactions, exchange, zero-point oscillations, inverse beta decay and chemical composition. Based on their calculations, a WD with a carbon core has a limiting mass of close to $1.40M_{\odot}$ \cite[tab.1b]{Hamada1961}. 
Most of these corrections are also included in our evolution code, which uses the (finite temperature) equation of state of \cite{Iben1992}. 
It is therefore not surprising that the instability occurred while the WD mass was still close to the initial assumed mass of $\sim1.40M_{\odot}$.

We have already demonstrated that WDs accumulate mass at high accretion rates \cite[]{Yaron2005}. We sampled the behavior of less massive WDs as well, in order to understand the manner in which a less massive WD can possibly accumulate the required additional mass to reach the Chandrasekhar limit. We ran models of $1.0$, $1.2$, $1.32$, $1.34$, $1.36$ and $1.38M_{\odot}$ for an accretion rate of $5\times 10^{-7}M_{\odot}yr^{-1}$, examining the differences in the net accreted mass per cycle and the cycle duration. The results of this investigation show that as the WD mass grows, the net accreted mass decreases and the cycle duration shortens. The drastic difference in cycle duration for different masses, accreting at identical rates, is demonstrated in Fig.\ref{fig:P(Mdot)}. For example, a $1.4M_\odot$ WD accreting at a rate of $5\times 10^{-7}M_\odot yr^{-1}$ has a cycle duration of $\sim 50$ days while for the same accretion rate the cycle duration for a $1.2M_\odot$ is $\sim 3.3$ years.\par

Dividing the net accreted mass of a single cycle by the cycle duration yields the \textit{effective} accretion rate $\dot{M}_{\rm eff}(M)$ or growth rate of the WD mass as a function of the WD mass $M$. For the range of the WD masses we have explored, $\dot{M}_{\rm eff}(M)$ is almost constant, varying from $2.27\times 10^{-7}M_{\odot}yr^{-1}$ for a $1.0M_{\odot}$ to $2.65\times 10^{-7}M_{\odot}yr^{-1}$ for a $1.4M_{\odot}$.  
To obtain the time required for a WD of given initial mass $M_0$ to reach the Chandrasekhar limit, $\tau_{\rm SN}(M_0)$, we use 
\begin{equation}
\tau_{\rm SN}(M_0) = \int_{M_0}^{M_{\rm Ch}}\frac{dM}{\dot{M}_{\rm eff}(M)}.
\end{equation}
For example, a $1.0M_\odot$ WD, accreting at a constant rate $\dot{M}=5\times 10^{-7}M_{\odot}yr^{-1}$, will reach the Chandrasekhar limiting mass in $\sim 2\times 10^6$ years, well within the evolutionary time scale of a symbiotic binary system.

We can now use $\tau_{\rm SN}$ to determine the minimum donor mass ($M_{\rm s}$) required to allow a WD of a given initial mass $M_0$ to reach $M_{\rm Ch}$. The donor loses mass over $\tau_{\rm SN}$ years, at a given rate $\dot{M}$ during the accretion phase of each cycle. Thus, the effective rate of mass loss by the secondary is given by the accretion rate multiplied by a factor $0<\alpha<1$, which denotes the ratio between the accretion time of a cycle and the total cycle time. Hence the lower limit for the secondary mass is given by 
\begin{equation}
M_{\rm s}(M_0)>\tau_{\rm SN}(M_0)\times \dot{M}\times \alpha (M_0)
\end{equation}
Using the same example of a $1.0M_\odot$ WD, accreting at $5\times 10^{-7}M_\odot yr^{-1}$, the donor must be at least $\approx 0.4M_{\odot}$, which is consistent with nova theory.     
In further studies we intend to broaden this investigation over the relevant accretion rate limits we have found here, in order to determine a lower limit for $M_0$ that may lead to a SN explosion.

\subsection{Three Candidate Pre-SN}\label{:conclusions}
\begin{figure}
\begin{center}
{\includegraphics[viewport = 70 10 745 580,clip,width=0.99\columnwidth] {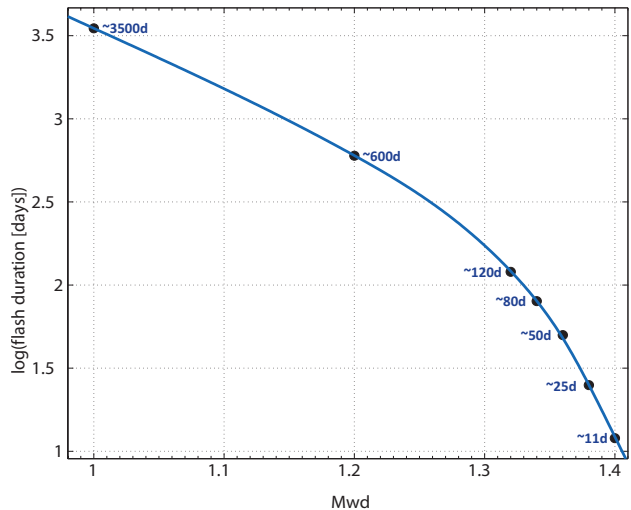}}
\caption{\label{fig:FlashDuration} Flash duration as a function of the initial WD mass.} 
\end{center}
\end{figure}

The recurrence time of a nova (as given in Fig.\ref{fig:P(Mdot)}) can be very helpful in identifying and estimating the mass and accretion rate of observed novae. Combined with the flash duration (i.e. the duration of the nova or nova-like explosion, often referred to as the {\textquotedblleft}optical{\textquotedblright} or {\textquotedblleft}low{\textquotedblright} state of the nova) should substantially narrow down the possibilities.  
The flash duration, presented in Fig.\ref{fig:FlashDuration}, is strongly dependent on the WD's mass and is shorter for a more massive WD, e.g., the flash duration for a $1.4M_\odot$ WD accreting at a rate of $5\times 10^{-7}M_\odot yr^{-1}$ is $\sim 11$ days while for a $1.2M_\odot$, accreting at the same rate, the flash duration is $\sim 600$ days.  

A classical nova eruption is usually first detected in the optical. As the optical decline begins, and the nova envelope contracts, bolometric corrections shift the energy output to the UV and X-ray passbands. Recurrent novae have alternating high-low states. Our models show novae recurring every few months to a few years. This should be observed as optical, UV and X-ray emission being detected on and off on a short time scale, and indeed there are reported cases that fit this description. \cite{Simon2005} and \citeauthor{Hachisu2012} (\citeyear[fig.5a]{Hachisu2012}) reported this variability, on a time scale of 180 to 420 days, for the super-soft X-ray source V Sge, which is believed to be a system with a very high mass transfer rate onto a WD \cite[]{Greiner1998,Simon1999}. They have suggested a state of accretion wind evolution as a possible process to explain the high accretion rate. 
The same type of variability has been reported for the super-soft X-ray binary RX J0513.9-6951 \cite[fig.1]{Burwitz2008}; \cite[fig.4a]{Hachisu2012}; \cite[fig.2]{Rajoelimanana2013} where the high state, lasting $\sim130-230$ days, is alternating with the low state, lasting $\sim30$ days.
The estimated mass accretion rate based on a parametrized model ranges from $7.9\times 10^{-8}-1.2\times 10^{-7}M_{\odot}yr^{-1}$ \cite[]{Kahabka1995} and up \cite[]{Southwell1996}. \cite{Southwell1996} and \cite{Pakull1993} concluded that the cause of the X-ray high state is most likely a WD's contracting atmosphere. 
All these estimates rely on the assumption of steady state burning. \cite{Wolf2013} define steady state burning as stable when an increase in temperature causes the cooling rate to increase more than the energy generation rate. In their models, the lower limit for steady state burning is $3.8\times 10^{-7}M_{\odot}yr^{-1}$ for a $1.34M_{\odot}$ WD and lower for less massive WDs. Their steady burning models are much like the extremely high accretion rate models from \cite{Yaron2005} that do not lose mass during flashes. 

\cite{Newsham2013} reported epochs of steady burning for all three of their models ($0.7$, $1.0$ and $1.35M_\odot$) accreting at high rates, which eventually either gives way to recurrent hydrogen flashes, transforms into a red giant or is interrupted by helium flashes depending on the mass and accretion rate. However, despite the mass loss due to the flashes, all of their models continue to grow in mass.

Our detailed evolutionary calculations, as well as those of \cite{Starrfield2012} and \cite{Idan2013} demonstrate that a state of steady hydrogen burning \textit{does not occur}. 
Cal 83 was observed with alternating optical/X-ray states as well, repeating every $\sim450$ days, spending about $\sim250$ days in the high state and $\sim200$ days in the low state \cite[fig.1]{Rajoelimanana2013}. The mass accretion model developed by \cite{vanden1992} requires an accretion rate of $\sim10^{-7}M_{\odot}yr^{-1}$. Using a $450$ day cycle in our power law (Fig.\ref{fig:P(Mdot)} and Eq.\ref{eqPMdot}) yields the same rate; $9.18\times 10^{-8}M_{\odot}yr^{-1}$. 
The model by \cite{Kahabka1995} yields a similar accretion rate of $1.2-3.3\times 10^{-7}M_{\odot}yr^{-1}$ and an extremely massive WD of $1.32-1.38M_{\odot}$. 
\cite{Greiner2002} reported shorter off-states for Cal 83 of $\sim50-120$ days.  \cite{Lanz2005} deduced from their best model fit, for a 50 day off-state, that the best match for Cal 83 would be a WD of mass $1.3\pm 0.3M_{\odot}$ and they maintain that Cal 83 is a very likely candidate for a future SNIa event. 
\cite{Rajoelimanana2013} argue that the orbital periods of Cal 83 and RX J0513.9-6951 (1.05 and 0.76 days) are too long to be {\textquotedblleft}{normal}{\textquotedblright} dwarf novae and in the absence of any other explanation are expected to have steady shell burning with periodic expansion and contraction of the WD's envelope. Expansion and contraction of the envelope do indeed occur, but we predict that the behavior of Cal 83 and RX J0513.9-6951 are the result of \textit{non-steady burning}.  

\subsection{Helium Flashes}\label{:Helium}
\cite{Jose1993} and \cite{Idan2013} show that helium flashes occur periodically on massive hydrogen accreting WDs every few tens of hydrogen flashes, depending on the WD mass and the accretion rate. In the course of a helium flash, a portion of the helium envelope is ejected. \par
\cite{Idan2013} modelled extremely high accretion rates where the entire helium shell was ejected over the course of a few consecutive flashes; however, they did not study the range of accretion rates investigated here.\par  
\cite{Newsham2013} found helium flashes for a WD mass of $1.35M_\odot$ accreting at rates higher than $5\times 10^{-7}M_\odot yr^{-1}$, but the velocity of the ejected envelope barely reached $3\%$ of the escape velocity, meaning the envelope will eventually fall back on to the WD.\par
\cite{Kato1999, Kato2004} calculated the mass accumulation efficiency during helium shell flashes on WDs of masses ranging from $0.6$ to $1.35M_\odot$. The value they obtained for the efficiency expresses the fraction of the envelope that undergoes nuclear burning into C+O (due to the high envelope temperatures during the helium flash) while the rest of the helium envelope is ejected. Their efficiency value is larger than 0.5 for their entire range of WD masses that gain helium at rates higher than $2\times 10^{-7}M_\odot yr^{-1}$, and it reaches as high as 1 for higher rates. The lowest value they obtained was 0.34 for a $1.2M_\odot$ WD gaining helium at a rate of $~8.7\times 10^{-8}_\odot yr^{-1}$. This means that periodic helium flashes will not prevent the WD from growing. In fact, the high temperatures during the helium flash will accelerate the nuclear burning of helium into carbon and oxygen, thus contributing to the growth of the WDs core. The affect of the helium flashes on results will be regarding the time frame; e.g., if the efficiency is 0.5 it will take twice as long for the WD to reach the Chandrasekhar limit. It is these boundaries, among others, that we intend to investigate in further studies.\par
We note, that in our models, no helium flashes developed. This is simply because the $1.4M_\odot$ WD reached the SNIa explosion before accumulating a sufficient amount of helium to ignite and trigger a helium flash.

\section{Conclusions}\label{:concl}
Can a WD with a hydrogen-rich donor in a close binary system somehow accrete sufficient mass to reach the Chandrasekhar limit and undergo a SNIa explosion? We have, for the first time, demonstrated via detailed numerical simulations that,
under the right conditions, it is indeed possible. A $1.4M_{\odot}$ WD accreting at a constant rate within the limits of $0.3$ to $6\times10^{-7}M_{\odot}yr^{-1}$, while undergoing periodic nova eruptions, eventually will experience a SNIa explosion. WDs in the range of at least $1.0-1.4M_\odot$ accreting at $5\times 10^{-7}M_\odot yr^{-1}$ will do so as well. We note that in reality, as the system evolves, the accretion rate will change due to donor bloating and angular momentum loss. We also note that, starting from a less massive WD, a significantly larger number of outbursts would precede the explosion.
In further research we intend to evolve the binary system as a whole in order to determine how wide the window of opportunity is for SNIa in evolving binary systems.

We have also found that the short-period outbursts preceding a SN explosion have observable signatures in UV and X-ray bands, which may serve to detect possible SNIa progenitors. We have found a few such examples among X-ray sources and intend to search for more.

Ending with a prediction, our models show that each cycle ejects an approximate mass of a few $10^{-8}M_{\odot}$. About $40\%$ of this is helium, $\sim2.5\%$ are heavy elements and the rest --- more than $57\%$ --- is hydrogen. Multiplying this by tens of thousands of cycles yields a total of a few $10^{-4}M_{\odot}$ expanding away from the WD. When the SN occurs, this optically thin expanding nebula will be compressed and eventually might become visible, leading to SNIa displaying hydrogen spectral lines, such as SN PTF 11kx \cite[]{Dilday2012}.

\section*{Acknowledgments} 
This work was supported by Grant No.2010220 of the United States --- Israel Binational Science Foundation and by the Ministry of Science, Technology, and Space, Israel.

\end{document}